%% file: socialsense.tex
\pdfoutput=1
\documentclass[sigconf]{acmart}
\usepackage{enumitem}

\usepackage{algorithm}
\usepackage{algpseudocode}

\begin{document}

%\title{The Fog Makes Sense: Enabling Social Sensing Services With Limited Internet Connectivity}
\title[Enabling Social Sensing Services With Limited Internet Connectivity]{The Fog Makes Sense: Enabling Social Sensing Services With Limited Internet Connectivity}

\titlenote{(c) Authors 2017. This is the author's version of the work. It is posted here for your personal use. Not for redistribution. The definitive version is published in the Proceedings of the 2nd International Workshop on Social Sensing, SocialSens'17, http://dx.doi.org/10.1145/3055601.3055614 }
%\subtitle{Extended Abstract}
%\subtitlenote{The ful	l version of the author's guide is available as
%  \texttt{acmart.pdf} document}

\author{Ruben Mayer}
\affiliation{%
  \department{Institute for Parallel and Distributed Systems}
  \institution{University of Stuttgart}
  \city{Stuttgart, Germany} 
}
\email{ruben.mayer@ipvs.uni-stuttgart.de}
\author{Harshit Gupta, Enrique Saurez}
\author{Umakishore Ramachandran}
 \affiliation{%
   \institution{Georgia Institute of Technology}
   \city{Atlanta} 
   \state{Georgia, USA} 
 }
\email{harshitg,esaurez,rama @gatech.edu}

\begin{abstract}
Social sensing services use humans as  sensor carriers, sensor operators and sensors themselves in order to provide situation-awareness to applications. This promises to provide a multitude of benefits to the users, for example in the management of natural disasters or in community empowerment. However, current social sensing services depend on Internet connectivity 
%of the social sensors and of the service users, as 
since the services are deployed on central Cloud platforms. In many circumstances, Internet connectivity is constrained, for instance when a natural disaster causes Internet outages or when people do not have Internet access due to economical reasons. In this paper, we propose the emerging Fog Computing infrastructure to become a key-enabler of social sensing services in situations of constrained Internet connectivity. To this end, we develop a generic architecture and API of Fog-enabled social sensing services. We exemplify the usage of the proposed social sensing architecture on a number of concrete use cases from two different scenarios.
\end{abstract}

\copyrightyear{2017}
\setcopyright{acmlicensed}
\acmConference[SocialSens'17]{The 2nd International Workshop on Social Sensing}{April 21 2017}{Pittsburgh, PA, USA}
\acmYear{2017}
\acmISBN{978-1-4503-4977-2/17/04}
\acmPrice{\$15.00}
\acmDOI{http://dx.doi.org/10.1145/3055601.3055614}




% We no longer use \terms command


\maketitle

\input{content/introduction.tex}
\input{content/architecture.tex}

\input{content/scenarios.tex}
\input{content/technical.tex}
\input{content/relatedwork.tex}
\input{content/conclusion.tex}

%\vspace{-0.2cm}
\bibliographystyle{ACM-Reference-Format}
\bibliography{socialsense}

\end{document}

%% file: content/introduction.tex
\section{Social Sensing}
\par Situation-aware applications use data streams from sensors to provide useful services to users or other applications. With the proliferation of sensors deployed in the surrounding world, e.g., through the Internet of Things, the potential of such applications is reaching new dimensions. Recently, research focus has been expanded from traditional fixed sensor deployments 
%e.g., temperature sensors installed in a building, 
toward \emph{social sensing} \cite{aggarwal2013social}. This comprises passive sensors provided by human carriers in Smart Phones, active human sensor operators taking pictures or videos and even humans operating as sensors themselves, e.g., providing live information in tweets and postings.
Recently, new applications have been proposed which use the social sensing infrastructure to infer situations that are not detectable from traditional sensors.
\subsection{Application Fields}
\par An important application field of social sensing is in helping people to deal with natural disasters. There are applications that help in finding friends and family in the aftermath of a natural disaster 
% or another incident in the chaos 
\cite{6081815}. Furthermore, social media can provide access to relevant and timely information to individuals in affected regions \cite{Simon2015609}. 
Providing real-time information to disaster-affected people about the situation in the area can help them take mitigative actions, for instance moving contents located in a flood-prone ground floor to upper floor \cite{allaire2016disaster} to reduce the loss caused by the disaster. Social media has been an effective way of sharing this sort of crowd-sourced information and can be more accurate and meaningful than government predictions.
Many proposals envision disaster-stricken people to perform social sensing tasks, like providing information about the level of inundation of roads in the event of a flood or tsunami. Such unstructured information would be mined by a social sensing application to extract relevant details and create a map of the affected area with important information. These maps can be used by government agencies to perform rescue operations \cite{eilander2016harvesting}. %Social sensing users 
Users can upload pictures of people with them, and social sensing applications apply face recognition algorithms on the pictures and let the friends and family of detected individuals know that they are safe. 
\par In rural or economically under-served regions, social sensing helps in understanding socioeconomic processes  \cite{doi:10.1080/00045608.2015.1018773} which can empower communities to better utilize their social capital%
\footnote{Social capital refers to the features of social organizations that facilitate coordination and cooperation for mutual benefit.} and
enable self-organized governance. Public transportation in such regions leave much to be desired due to lack of consistency in schedules and infrastrual support, forcing passengers to wait for long periods of time. In well-served communities, infrastructual support (e.g., kiosks at bus stops operating on GPS data) provide timely information for the passengers. 
%The lack of information related to the infrastructure restricts the benefits the people could derive from it. 
Social sensing services in such under-served regions could help gather information, e.g., when the bus is going to arrive and share with others even in the absence of infrastructural support. 
%\par Kishore's example about the latency tolerant sensing. The social aspect is that people might harvest the data with their private devices and transmit them to the Internet when available.

\subsection{Challenges}
\par While the discussed applications are very effective in utilizing social sensing information, they rely on Internet connectivity of the social sensors, the situation inference applications, and the users that are interested in the detected situations. This is mainly the case because the social sensing service is hosted in a central (cloud) data center.
\par However, Internet connectivity cannot be taken for granted on any of the layers of a social sensing application. Internet outages can affect large areas in case of emergencies, natural disasters, or hacker attacks on the Internet infrastructure \cite{Manoj:2007:CCE:1226736.1226765}. Furthermore, rural regions might not be connected to the Internet at all, or the inhabitants of a rural or an under-served urban region cannot afford Internet connectivity for economical reasons. Social sensing applications can be of a huge benefit in exactly such situations and circumstances.  All of those benefits are tightly coupled to the Internet connectivity; without the Internet, social sensing services are not available. 
\par In recent years, a new trend has emerged in computing infrastructures that can help in overcoming the Internet dependency of social sensing services. \emph{Fog Computing}, also known as Edge Computing, is the approach of adding computational resources toward the edge of the Internet \cite{bonomi2014fog}. While it was initially intended to improve network latency between sensors, applications, and users  \cite{Hong:2013:MFP:2491266.2491270}, we propose Fog Computing to become a central enabler of decentralized, local social sensing services that can also operate when Internet connectivity is constrained. This way, social sensing services can become more robust to Internet outages. Furthermore, communities that did not benefit from the first wave of cloud-based social sensing services can leapfrog those and directly use Fog-based services.  %Democratizing the benefits of cloud computing to all the communities that were isolated from the first wave of applications. 

% \begin{itemize}
% \item \begin{figure}
% \item     \centering
% \item     \includegraphics[width=0.75\linewidth]{graphics/architecture}
% \item     \vspace{-0.25cm}
% \item     \caption{System Model}
% \item     \label{fig:architecture}
% \item     \vspace{-0.3cm}
% \item \end{figure}
% \end{itemize}

\begin{figure}
     \centering
     \includegraphics[width=0.65\linewidth]{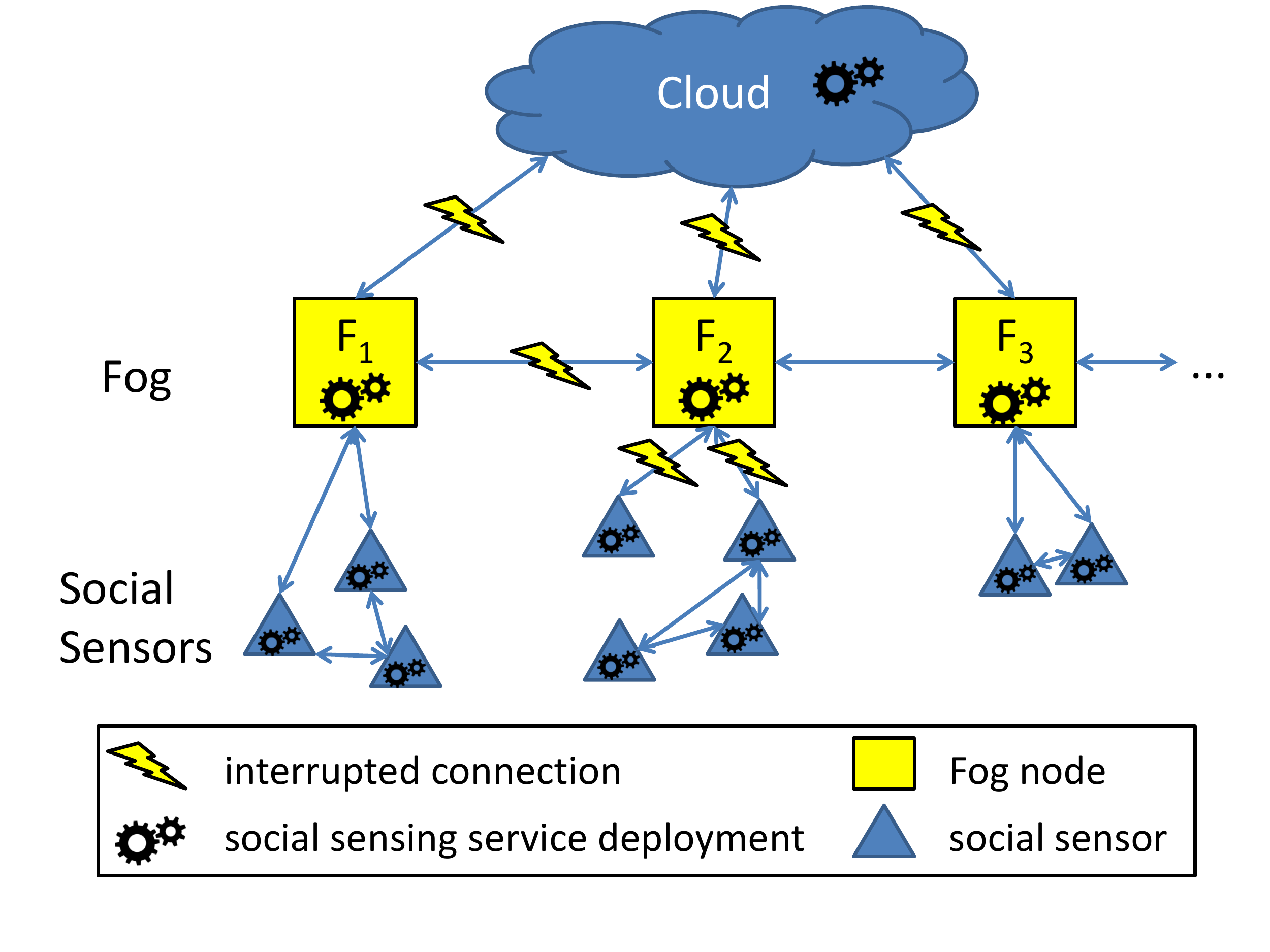}
     \vspace{-0.25cm}
     \caption{System Model}
     \label{fig:architecture}
     \vspace{-0.5cm}
 \end{figure}

\par However, today's social sensing services are not capable of using the Fog infrastructure to provide local services when Internet connectivity is impaired. It is not enough to just run a centralized social sensing service on a number of Fog nodes in parallel. Instead, the social sensing service has to become a \emph{distributed service} capable of discovering available Fog nodes and building a network that aggregates and shares information between social sensors that are connected to different Fog nodes. In this regard, it needs to be able to deal with the volatile nature of Fog and sensor connectivity. To this end, the architecture of social sensing services needs to be adapted to fully utilize the opportunities of the Fog infrastructure.

\subsection{Outline}
\par In this paper, we give an overview of evolving Fog-based computing infrastructures. Based on that, we propose a generic architecture for Fog-based social sensing services. Using two concrete case studies, we demonstrate how existing cloud-based social sensing services can be adapted to use the Fog-based architecture. We conclude that utilizing Fog-based computing architectures is a promising path to more robustness and democratization of social sensing services.

%% file: content/architecture.tex
\section{Fog-based Social Sensing Architecture}

\subsection{Overview}
In the following, we give an overview of the emerging Fog Computing architecture. We point out that the Fog infrastructure can be completely heterogeneous. Social sensing %services 
on Fog has to be able to cope with the heterogeneity provided in the available resources. %Different Fog nodes of a different type should be able to collaborate in providing a uniform service. % (i) Robust Fog nodes operating in bunkers with battery or Diesel generators. (ii) Mobile Fog nodes on drones, balloons, etc. that are deployed by emergency response. (iii) People providing the local connectivity, e.g., WiFi, of their battery-powered Mobile Phones or local computational power in their homes. 

\par Figure \ref{fig:architecture} shows a model of the Fog Computing architecture. On the top layer, the traditional Cloud data center is depicted, being deployed in the core of the network and only reachable via Internet connections. Such data centers are characterized by using standardized, off-the-shelf computing resources, and a virtualization layer that allows for an effective utilization of the resources and a pay-as-you-go business model.
In the middle layer, a number of heterogeneous Fog nodes are geographically distributed deployed at the edge of the network. This means, that Fog nodes can be locally reachable by connected devices nearby, even if the Internet is not available. On the bottom layer, geographically distributed social sensors are connected to their close-by Fog nodes, either directly or by using other social sensors as relays.
% Neighboring Fog nodes that are still inter-connected can share information, increasing the local range of the social sensing service

\begin{figure}
\centering
\begin{tabular}{|p{1.5cm}|p{2cm}|p{1.6cm}|l|}
\hline
Device & Computational Capabilities & Containers Supported? & Connectivity \\
\hline\hline
Routers & Low & Yes & WiFi, LAN \\
\hline
$\mu$Computers & Medium & Yes & WiFi, LAN \\
\hline
Drones & High & Yes & WiFi, 4G \\
\hline
%Smart Phone & Medium & No & WiFi, LAN \\
\end{tabular} 
    \vspace{-0.3cm}
    \caption{Overview of Fog Devices.}
    \label{fig:devices}
    \vspace{-0.5cm}
\end{figure}

As there is a heterogeneity of use cases for Fog computing, there are many different notions of a Fog node. In the following, we provide an overview of current proposals and products (cf. Figure \ref{fig:devices}). With the advent of computationally stronger network equipment, especially routers, it has been proposed that computations are already performed in the network. For instance,  Cisco offers their IOx platforms on hardened routers \cite{cisco} that are capable of performing data processing tasks. On a higher layer, mini-computers like Raspberry Pi have gained popularity, as they provide acceptable computation performance for a very low price. Additionally, the energy efficiency and miniaturization of those devices allow them to run in environments that were not specifically designed to host computers, i.e., outside of data centers. Mini-computers can even be deployed on drones \cite{7127659} and provide a completely new level of ``mobile computing''. A swarm of drones can build an ad-hoc network, a so-called Flying Ad-Hoc Network (FANET) \cite{Bekmezci20131254}, and this way provide Fog computing in an area that lacks any infrastructure. Generally, the deployment of Fog services can be facilitated by using recent lightweight container technology like Docker \cite{Bellavista:2017:FFC:3007748.3007777}.
\par The social sensors can be \emph{smart sensors} that perform the sensing, but also filtering and aggregation. In the scenarios described, typically the smart sensors would be connected to smart phones which have certain computational capabilities to do the filtering and aggregation. This reduces the communication overhead between social sensors and Fog nodes, and also reduces computational overhead on the Fog nodes.

\begin{figure}
    \centering
    \includegraphics[width=0.65\linewidth]{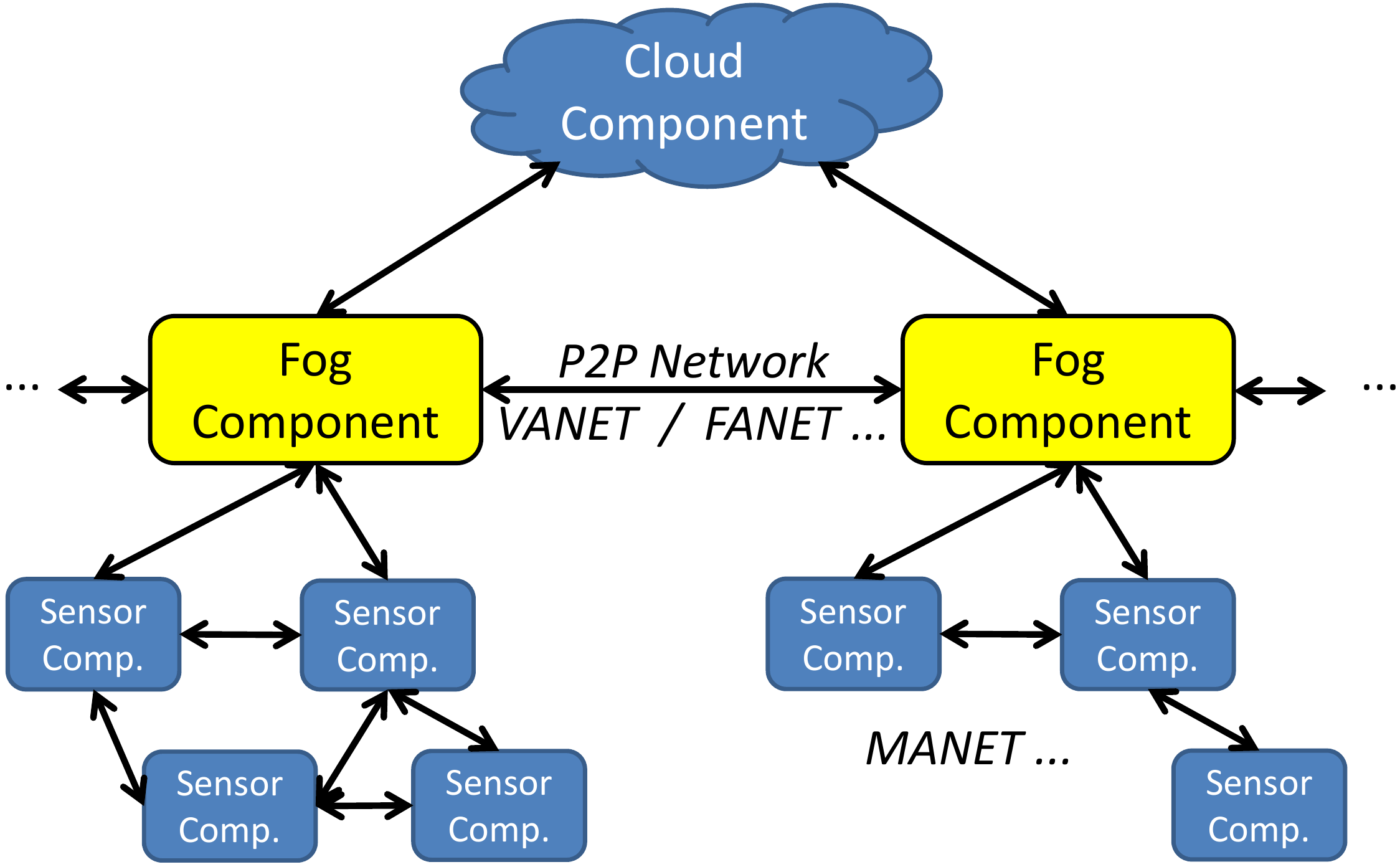}
    \vspace{-0.3cm}
    \caption{Generic Software Architecture.}
    \label{fig:software_architecture}
    \vspace{-0.5cm}
\end{figure}

\subsection{Fog-Enabled Social Sensing Services}
Here, we analyze how social sensing services can exploit the Fog infrastructure. They should be able to operate on local information provided on a single Fog node, but also capable of sharing information and collaborating with social sensing services running on neighboring Fog nodes that are reachable. Finally, if the Cloud is reachable, the social sensing services on the different Fog nodes should be able to share global information via the Cloud.
\par We propose a \emph{generic software architecture} for social sensing applications that is capable of exploiting the Fog infrastructure (cf. Figure \ref{fig:software_architecture}). It consists of three components: (i) A central management components placed in the Cloud infrastructure (the \emph{Cloud Component}), (ii) A data processing component placed in the Fog infrastructure (the \emph{Fog Component}) and (iii) a social sensing component deployed on the users' devices (the \emph{Sensor Component}). In the following, we detail the tasks of the components and provide an API on all the components for the developers of social sensing services.
\subsubsection{Cloud Component}
The Cloud Component is, first of all, responsible for the deployment and management of the Social Sensing Service artifacts (program code, meta-data, settings, etc.) on the Fog Components, when an Internet connection is available. This means that the Cloud Component sends software updates to the Fog Components---which can be forwarded to the Sensor Components from there---and also gathers status information from the Fog Components. This can, e.g., help in deciding where to deploy more Fog Components. Furthermore, the Cloud Component can obtain a global view of the sensed data, which can be useful for offline analysis, e.g., in the aftermath of a natural disaster.  
\subsubsection{Fog Component}
\par The Fog Component is responsible for querying reachable social sensors for data. Such data queries can be formulated in state-of-the-art Stream Processing or Complex Event Processing query languages, such as CQL \cite{Arasu:2006:CCQ:1146461.1146463} or Tesla \cite{Cugola:2010:TFD:1827418.1827427}. This allows for defining \emph{continuous queries} that employ windows, aggregation functions and event patterns on the data streams. This way, the Sensor Components can aggregate and filter the sensed data before transferring it to the Fog nodes. Besides traditional continuous queries, queries can also ask for one-time manual sensing, e.g., querying a person to take photos of a specific scene or provide feedback about the number of persons in the vicinity. 
\par The results of the sensing queries are further aggregated in the Fog Component. Data streams from different social sensors need to be correlated so that an overall picture of the situation can be derived \cite{7024105}. Furthermore, the comparison of information from different sources can improve the information quality \cite{Simon2015609}.
\par Besides querying and aggregating sensor data from the social sensors in its vicinity, the Fog Component manages the sharing of information between different Fog nodes that are connected to each other over the network. This can be achieved with a peer-to-peer based communication network. Furthermore, if an Internet connection is available at a Fog node, local data can be streamed to the cloud for further analysis. Note that in latency-tolerant applications, mobile Fog nodes can serve as ``data mules'' that collect sensor data from an area that is disconnected from the Internet, transport that data to an area where the Fog node has Internet connectivity and forward the data to the cloud from there. This can be useful for retrieving sensor data from remote areas.

\subsubsection{Sensor Component}
\par The Sensor Component gathers the raw social sensing data, performs the queries from the Fog Component on that data and returns the results back. 
\par It should be noted that not all Sensor Components might be able to directly connect to a Fog Component. This can be due to their physical distance to the next Fog Component, or due to device limitations (e.g., supporting the communication requirements of the Fog Components). For instance, if the Fog Components all require 4G connectivity, some of the Sensor Components might not be able to directly connect to any Fog Components at all. Still, such Sensor Components could connect to other Sensor Components in their proximity, for instance, using WiFi networking. Then, a Sensor Component with direct access to a Fog Component serves as a \emph{communication relay} between the other Sensor Components and the Fog. Such a network can, for instance, be realized with methods from Mobile Ad-Hoc Networks (MANETs). A similar idea was presented by Yusuf, et al \cite{lateef2012} with the micrograph middleware. It shows how to handle discovery and manage these distributed and isolated communities for social networks.
\par Note that as the Sensor Components can be disconnected from the Fog  at any time, e.g., because the Fog Component goes down, continuous queries on the Sensor Components should be \emph{soft state}, i.e., employ a time-out mechanism; when the connection to the Fog layer is interrupted for a long time, the sensing is stopped to save energy on the social sensing devices.
\subsubsection{Programming API}

\begin{figure}
\centering
\begin{tabular}{|p{1.2cm}|p{3.1cm}|l|}
\hline
Compo-nent & Primitive & Parameters  \\
\hline\hline
Cloud  & \texttt{publish} & topic, message \\
 \hline
 & \texttt{subscribe} & topic, message handler \\
\hline \hline
Fog & \texttt{on\_sensor\_connection} & sensor node \\
\hline
 & \texttt{on\_fog\_connection} & fog node \\
\hline
 & \texttt{on\_cloud\_connection} & cloud node \\
\hline
 & \texttt{query\_specific\_sensor} & sensor node, query \\
\hline
 & \texttt{query\_all\_sensors} & no. of answers, query \\
\hline
%  & \texttt{send\_to\_peer} & peer node, message \\
% \hline
%  & \texttt{send\_to\_cloud} & cloud node, message \\
% \hline
%  & \texttt{on\_sensor\_msg} & sensor node, message \\
% \hline
%  & \texttt{on\_fog\_msg} & fog node, message \\
% \hline
%  & \texttt{on\_cloud\_msg} & cloud node, message \\
% \hline
 & \texttt{publish} & topic, message \\
 \hline
 & \texttt{subscribe} & topic, message handler \\
\hline \hline
% Sensing & query\_fog & fog node, query \\
% \hline
Sensor & on\_specific\_query & fog node, query \\
\hline
 & on\_general\_query & fog node, query \\
 \hline
 & \texttt{publish} & topic, message \\
 \hline
 & \texttt{subscribe} & topic, message handler \\
\hline
%Smart Phone & Medium & No & WiFi, LAN \\
\end{tabular} 
    \vspace{-0.3cm}
    \caption{Programming API for Social Sensing Services.}
    \label{fig:api}
    \vspace{-0.3cm}
\end{figure}

\par We propose basic APIs for the Cloud, Fog and Sensor Components (cf. Figure \ref{fig:api}). Designers of social sensing services can use those APIs in order to design their system. These APIs are based on a \emph{reliable and delay-tolerant publish-subscribe middleware} \cite{skjegstad2012mist}, which allows information dissemination between nodes in a dynamic network, which is exactly the case with social sensing. %In the following, we explain the proposed primitives in more detail.

\paragraph*{API of the Cloud Component}
\begin{description}[leftmargin=0pt,align=left]
    \item[\texttt{publish}:] Cloud component can publish to a specific topic, which serves as the medium of sending message to fog nodes. 
    \item[\texttt{subscribe}:] Cloud Component can subscribe to messages published to the specified topic by a connected Fog Component, and specify a handler function to be called when such a message arrives.
\end{description}

\paragraph*{API of the Fog Component} 
\begin{description}[leftmargin=0pt,align=left]
    \item[\texttt{on\_sensor\_connection:}] Called when a sensor connects to the Fog node. The corresponding Sensor Component is registered in the Fog Component, so that future queries can be directed to that Sensor Component. In the registration process, context information of the sensor is exchanged, e.g., the location, and whether other neighboring Sensor Components are reachable by the connecting Sensor Component.
%     \item[\texttt{on\_fog\_connection}] Called when a Fog node connects to another Fog node, accompanied by registration of peer on both nodes and exchange of context information, e.g., the location and further Fog Components and Sensor Components that are reachable by the peer. %The corresponding Fog Component is registered in the peer-to-peer network. In the registration process, context information of the Fog node is exchanged, e.g., the location and further Fog Components and Sensor Components that are reachable by the connecting Fog Component.
%     \item[\texttt{on\_cloud\_connection}] Called when a Fog Component can connect to the cloud data center. The Fog Component can now exchange data and queries with the cloud.%now exchange data with the cloud, for instance, send data about the current situation, and receive queries from the cloud to investigate specific situations.
    \item[\texttt{query\_specific\_sensor:}] Primitive for querying a specific Sensor Component. The query can be a continuous query formulated in a Stream Processing or a Complex Event Processing query language, but also a one-time query that requires manual action by the social sensor, e.g., taking a photo of a specific scene or reporting how many persons are in the social sensor's vicinity.
    \item[\texttt{query\_all\_sensors:}] Primitive for broadcasting a query to sensor components, expecting to get a specific number of query responses, without needing to specify which specific Sensor Components should receive that query. This is useful to achieve a high-fidelity view of the situation by aggregating multiple responses.
    %does not want to rely on single Sensor Components, but aggregate a larger number of answers to get a higher quality view of the situation. 
%     \item[\texttt{send\_to\_peer}] The Fog Component can send messages to any of its peer Fog Components. This way, it can share information about the local situation, but also request to get further information from the neighboring peers. When Fog Components are connected, this way, they can get an aggregated view of a larger area.
%     \item[\texttt{send\_to\_cloud}] The Fog Component can send messages to the cloud if it is connected. This way, it can share information on the local situation to allow for further global analysis in the cloud.
     \item[\texttt{publish:}] The Fog Component can publish a message to a specific topic to convey information to peer fog nodes or the cloud service.
     \item[\texttt{subscribe:}] The Fog Component can subscribe to messages published to the specified topic, and specify a handler function to be called when such a message arrives.

%     \item[\texttt{on\_sensor\_msg}] The Fog Component can receive messages from its connected Sensor Components. Those are typical answers to the installed queries.
%     \item[\texttt{on\_fog\_msg}] The Fog Component can receive messages from its connected peer Fog Components. See \texttt{send\_to\_peer}.
%     \item[\texttt{on\_cloud\_msg}] The Fog Component can receive messages from the cloud data center if connected to the Internet. For instance, the cloud can send queries to the Fog Component, which itself can query the necessary data from the connected Sensor Components.
\end{description}

\paragraph*{API of the Sensor Component}
\begin{description}[leftmargin=0pt,align=left]
%     \item[\texttt{query\_fog}] The Sensing Component sends a query to the local Fog Component it is connected to. The Fog Component handles the query accordingly, e.g., formulating queries to its local Sensing Components or forwarding the query to peer Fog Components. When the local Fog Component has a result to that query, it sends it to the querying Sensor Component.
    \item[\texttt{on\_specific\_query:}] Handler called when Sensor Component receives a query. If it is a continuous query, it installs the query, processes the sensed data accordingly and publishes results to the given topic. If it is a one-time query that requires manual action by the social sensor, it starts the appropriate routine, e.g., a message on the display of a smart phone.
    \item[\texttt{on\_general\_query:}] The Sensor Component receives a query that requires a given number of answers. The Sensor Component uses broadcast or multicast protocols, e.g., gossiping, to disseminate the query to an appropriate number of other Sensor Components. 
    \item[\texttt{publish:}] Sensor components can publish to a specific topic, which serves as the medium of returning query results to fog nodes. 
    \item[\texttt{subscribe:}] Sensor Component can subscribe to messages published to the specified topic by a connected Fog Component, and specify a handler function to be called when such a message arrives.
\end{description}

To create a Fog-enabled social sensing service, the Cloud Component, Fog Component and  Sensor Component can implement the proposed API.  In the next section, we give concrete examples of how the proposed API can be used in practice.

%% file: content/scenarios.tex
\begin{figure}
    \centering
    \includegraphics[width=0.9\linewidth]{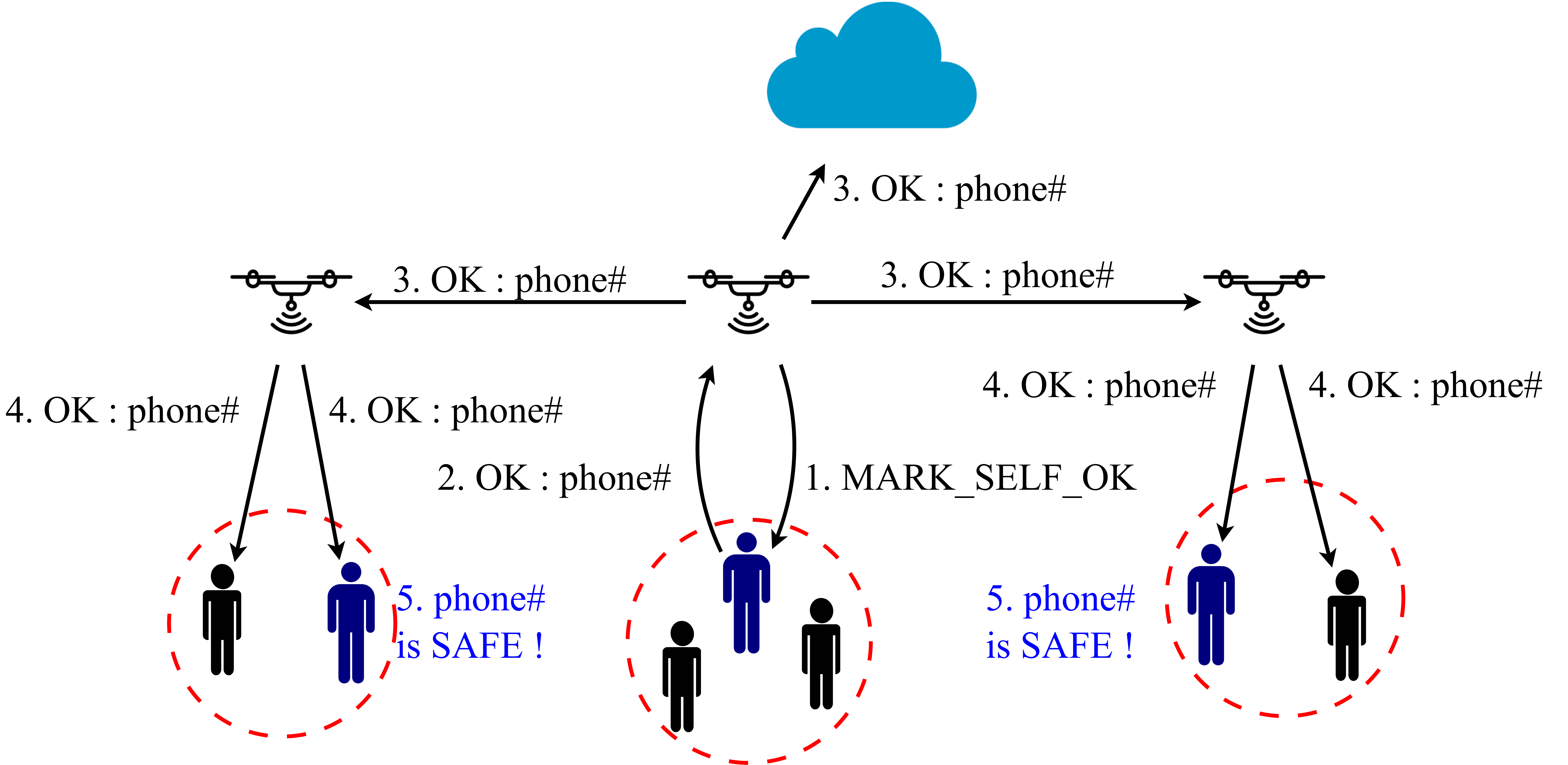}
    \vspace{-0.3cm}
    \caption{Schematic of Family Safety application (members of a family are distinguished by using blue color).}
    \label{fig:disaster_pubsub}
    \vspace{-0.3cm}
\end{figure}

\section{Case Studies}
In this section, we %analyze on a number 
discuss a couple of use cases to illustrate how social sensing services can be adapted to the proposed Fog infrastructure. We show that existing services could benefit from the deployment in the Fog by providing local services to their users in the face of Internet connectivity problems. 
\subsection{Social Sensing Services in Natural Disasters}
Social sensing services deployed on the Fog can help to gather and disseminate local knowledge among the affected people.
 Owing to the relatively local nature of the information pertaining to a disaster-prone area, Fog Computing is destined for providing the required connectivity to affected people and emergency response teams so that they can help mitigate the adverse effects.
\subsubsection{Checking the Safety Status of Family Members} Recall the use case that a person wants to check whether his family members in a disaster-stricken area are OK. We have discussed that the Internet connectivity can be interrupted, so that his family members cannot access cloud services like Facebook Safety Check\footnote{https://www.facebook.com/about/safetycheck/}. In such a situation, a Fog-enabled Safety Check Service can enable disaster-struck people to ascertain the safety of their family members, as shown in Figure \ref{fig:disaster_pubsub}. 
% \par With a Fog-enabled safety check service, he can query his local Fog Component, e.g., with the phone numbers of his relatives. To this end, the Sensing Component sends a query to its local Fog Component (using the primitive \texttt{query\_fog(local\_fog, "safety\_check(list\_of\_telephone\_numbers)"}).

\par A user interested in knowing whether his family members are okay starts the Sensor Component of the safety check service on his smartphone. The Sensor Component (SC) collects the list of phone numbers belonging to the user's family members. 
\par Whenever a Fog Component (FC) is able to connect to a SC, it queries that SC with a specific query asking to mark itself OK. Upon receiving this query, the SC displays an alert on the smartphone's screen asking the user to mark himself OK. Then, the SC publishes the user's phone number to its local FC, so that the FC can know about this user's safe situation.
\par All FCs subscribe to the topic named \textit{OK}, to which both SCs and peer FCs publish. The pub/sub system is the medium of disseminating information of safe users over the area. When an FC receives a new \textit{safe} phone number, it publishes it on the  topic ``OK'' so that it may be received by SCs and peer FCs. SCs also subscribe to the topic \textit{OK} so that they receive information about the people who are safe. If an SC receives  a \textit{safe} phone number and it belongs to one of the user's family members, she is alerted of their safety. 
%{\bf: ***Kishore***: Not clear what 1, 2, 3, 4, 5 are...line numbers in the algorithms?} Answer from Ruben: They were meant to be the steps from Figure 5. Left the numbers out to avoid confusion, as there are both, a Figure and an Algorithm.

% creates one-time queries to the connected Sensing Components requiring to respond whether they are OK (using the primitive \texttt{query\_specific\_sensor(sensor\_no, "safety\_check"})). Suppose that 2 out of 3 family members are connected to the same Fog Component as the querying person is. The local Fog Component sends safety check queries to those Sensor Components directly. Over the peer-to-peer network to other Fog Components, it searches for the Fog Component the third queried person is connected to. It forwards the query also to that Fog Component (using the primitive \texttt{send\_to\_peer}). Eventually, all queried persons' Sensor Components receive the query and send the positive response. The responses are aggregated in the Fog Component and the result ``everybody is OK'' is returned to the querying person. The local connectivity of the Fog Components was enough to find all persons. A social sensing service deployed in a central cloud data center would not have been reachable by any of the Sensing Components.
\begin{algorithm}
\caption{Fog component: Family Safety application}
\begin{algorithmic}
\Procedure{Init}{}
\State subscribe(OK, $on\_ok\_callback$)
\State safeList $\leftarrow \{\}$ \Comment{People this node knows are OK}
\EndProcedure
\Procedure{on\_ok\_callback}{$p$}
\If{$p \notin safeList$}
\State publish($OK, p$) \Comment{publish phone\# to OK people topic}
\State $safeList \leftarrow safeList \cup \{p\}$
\EndIf
\EndProcedure
\Procedure{$on\_sensor\_connection$}{$sensor$}
\State $query\_specific\_sensor(sensor, MARK\_SELF\_OK)$
\EndProcedure
\end{algorithmic}
\end{algorithm}
\vspace{-0.5cm}
\begin{algorithm}
\caption{Sensor component: Family Safety application}
\begin{algorithmic}
\Procedure{Init}{}
\State subscribe(OK, $on\_ok\_callback$)
\State family $\leftarrow$ GetFamily()
\EndProcedure
\Procedure{on\_ok\_callback}{$phone\#$}
\If{$phone\# \in family$} \Comment{if recv. phone\# is family}
\State DISPLAY(f is SAFE)
\EndIf
\EndProcedure
\Procedure{$on\_specific\_query$}{$type$}
\If{$type == MARK\_SELF\_OK$}
\State DISPLAY(Mark yourself OK)
\If{user marked OK} \Comment{if user marks himself OK}
\State publish($OK, phone\#$)
\EndIf
\EndIf
\EndProcedure
\end{algorithmic}
\end{algorithm}
\subsubsection{Population Density Map for Emergency Response Teams} Suppose that an emergency response team wants to get information about the distribution of individuals in an area struck by a hurricane. To this end, it needs aggregated information about the sectors in which the persons are located. Based on this, the emergency response deduces where to go first to help or evacuate people. 
\par The emergency response team installs a continuous query on the Fog Component of the area they plan to go to, querying for detailed information of how many Sensor Components are connected in which area. The continuous query is installed on the Fog Component using the primitive  \texttt{query\_all\_sensors($\infty$, pos\_query)}. As a result, all connected Sensor Components report their position. This data is used for building a density map of persons in specific sectors, which is returned to the emergency response team.
%\par A key characteristic of such an application is its ad-hoc peer-to-peer mode of execution on fog devices. Owing to limited connectivity range of fog devices, multiple fog devices are deployed in a disaster-affected area - that share information with each other opportunistically - so as to connect people across the affected area. An opportunistic push of information to the cloud allows other bodies like government agencies or media know about the situation of the disaster's aftermath.

\subsection{Social Sensing Services in Economically Under-Served Regions}
\subsubsection{Modernizing Public Transportation}
\par Using Fog Computing, social sensing services can learn about the transportation infrastructure: density of users in a bus, the variability of the service, predicted time tables, and other metrics. When facing intermittent Internet connectivity, Fog nodes can provide local connectivity to the users. To this end, each bus carries a small Fog computing device. Passengers in the bus can connect with their smart phones as social sensors, so that the Fog Component on the bus receives sensor data from the smart phones. For instance, the Fog Component can query the destination of the passengers (using the \texttt{query\_all\_sensors($\infty$, destination\_query)} primitive), so that an optimal bus route is calculated. 
\par Buses can share information with each other when they are close enough to build a vehicular ad-hoc network (VANET) \cite{lochert2003routing}. To this end, they use the \texttt{publish} primitive. Reliable and delay-tolerant publish/subscribe  ensures that buses can also share information with the cloud when an Internet connection is available. The shared information can concern, e.g., road and traffic conditions, so that the bus driver can benefit from the knowledge available in other buses as well.
Passengers in the bus can connect to the Fog Component to receive information on the expected arrival time at their destination, by subscribing to the appropriate topic.   
% Passengers in the bus can connect to the Fog Component to receive information on the expected arrival time at their destination, using the  \texttt{query\_fog} primitive.   

\par If passengers do not have smart phones that can carry Sensor Components of the system, installed Sensor Components on-site can provide them a basic service. For instance, a ``call a bus'' button can be installed at a bus stop. If a passenger needs a bus ride, she can press the button; the social sensing service will contact buses that are close-by via the Fog infrastructure and adapt the route such that the next bus will visit the queried bus stop. Such simple sensors are cheap enough to be deployed in rural and economically under-served areas in a large number.
Additionally, the information obtained can be used by local governments to improve the service and to ask other authorities for services that adapt better to the communities that are being served.
\subsubsection{Sensing the Status of Infrastructure}
In remote regions, it is difficult to monitor installed infrastructure, such as street lighting and solar panels, for failures. Employing them with sensors that sense the infrastructure status promises to facilitate the monitoring. Such sensors are available for a very low price. However, in rural areas, there might not be Internet connectivity to read the sensed data remotely. Even if Internet connectivity is available, providing the sensors with an Internet connection increases their price, for acquisition as well as operation. 
\par To this end, the idea of ``data mules'' has been proposed. People or moving objects such as buses can serve as a data mule to collect sensor data and upload it to the Cloud as soon as Internet connectivity is available. This has been proposed in order to deliver emails to remote regions \cite{1319279}. However, it is not directly applicable to sensor data, which might have much larger scales. When there is a lot of data to be sensed, the storage of micro-computers could be over-utilized. Furthermore, the Internet connection, when available, could still have a very low bandwidth, so that the upload of Gigabytes of data is infeasible. Extending the simple data mule concept to a mobile Sensor Component can help to handle this issue. The Sensor Components execute aggregation and filtering operations, so that the amount of data transmitted is much lower.

%% file: content/technical.tex
\section{Technical Challenges}
\par The Fog infrastructure poses a large range of technical challenges on the implementation. For example, if the Fog nodes are installed on drones, different communication protocols are used and coupling between them is required. Additionally, the network protocols need to be latency-tolerant; each node needs to be able to queue messages until a connection is reestablished.
\par Handling geo-distributed resources is challenging. Part of the complexity is defining the type of algorithm to deploy on the nodes based on the available capabilities. This has to be added to the process of deploying applications to nodes with limited Internet connectivity and untrusted infrastructure. 
\par Common distributed systems issues also arise in the context of Fog social sensing. Fog resources might have lower availability and dependability than servers in cloud data-centers. One of the main challenges is that protocols and middleware need to be distributed and energy-efficient, e.g., discovering other peers and fog nodes without a central entity and with limited energy. Load balancing is another common issue. For example, the region of a disaster may require more resources such as networking and computing. How can the Fog infrastructure be organized to meet different resource demands? Mobility of Fog nodes could be used to dynamically balance the pressure on each Fog node.
\par There also exist social sensing specific challenges. Social sensing can bring disinformation and inaccuracy \cite{Simon2015609}. The identity of the users and correctness of the information cannot be guaranteed \cite{merchant2011integrating}.  These errors are not necessarily intended, but caused by mistakes and misunderstandings. Fog computing itself enhances the sharing of information within the region responsible for a given Fog node. However the question arises, is there more we can do to provide reliable information sharing? For example, an intuitive idea is to gather the information from different social sensors and eliminate outliers. A further question is how to route the information to the intended receivers. Simple flooding will lead to each user receiving too much information and bringing pressure to the network infrastructure. When the Fog nodes are mobile, this issue becomes more challenging due to non-deterministic connectivity.

%% file: content/relatedwork.tex
%\section{Related Work}
%Related work on Fog Computing.

%Related work on social sensing in the Cloud.

%% file: content/conclusion.tex
\section{Conclusion and Outlook}
In this paper, we have extended the vision of Fog Computing toward providing social sensing services in situations when Internet connectivity is limited. We have outlined the basic design principles of such a Fog-enabled social sensing service, and have proposed a generic API that social sensing services can employ in order to use the Fog infrastructure.
\par A Fog-based social sensing platform could help in the transition of existing cloud-based social sensing services to the Fog. Such a middleware needs to provide basic building blocks for the developers of a social sensing service, and also include mechanisms for deployment, communication management, and many other technical details. It is imaginable that cloud-based services could be transformed toward the Fog platform in a semi-automatic manner. This way, many of the existing useful cloud-based services could benefit from the emerging Fog infrastructure.

%% file: socialsense.bbl

\begin{thebibliography}{00}

%%% the Web address.

\ifx \showCODEN    \undefined \def \showCODEN     #1{\unskip}     \fi
\ifx \showDOI      \undefined \def \showDOI       #1{{\tt DOI:}\penalty0{#1}\ }
  \fi
\ifx \showISBNx    \undefined \def \showISBNx     #1{\unskip}     \fi
\ifx \showISBNxiii \undefined \def \showISBNxiii  #1{\unskip}     \fi
\ifx \showISSN     \undefined \def \showISSN      #1{\unskip}     \fi
\ifx \showLCCN     \undefined \def \showLCCN      #1{\unskip}     \fi
\ifx \shownote     \undefined \def \shownote      #1{#1}          \fi
\ifx \showarticletitle \undefined \def \showarticletitle #1{#1}   \fi
\ifx \showURL      \undefined \def \showURL       #1{#1}          \fi
% invisible to TeX
\providecommand\bibfield[2]{#2}
\providecommand\bibinfo[2]{#2}
\providecommand\natexlab[1]{#1}

\bibitem[\protect\citeauthoryear{Aggarwal and Abdelzaher}{Aggarwal and
  Abdelzaher}{2013}]%
        {aggarwal2013social}
\bibfield{author}{\bibinfo{person}{Charu~C Aggarwal} {and}
  \bibinfo{person}{Tarek Abdelzaher}.} \bibinfo{year}{2013}\natexlab{}.
\newblock \showarticletitle{Social sensing}.
\newblock In \bibinfo{booktitle}{{\em Managing and mining sensor data}}.
  Springer, \bibinfo{pages}{237--297}.
\newblock


\bibitem[\protect\citeauthoryear{Allaire}{Allaire}{2016}]%
        {allaire2016disaster}
\bibfield{author}{\bibinfo{person}{Maura~C Allaire}.}
  \bibinfo{year}{2016}\natexlab{}.
\newblock \showarticletitle{Disaster loss and social media: Can online
  information increase flood resilience?}
\newblock \bibinfo{journal}{{\em Water Resources Research\/}}
  \bibinfo{volume}{{52}, 9} (\bibinfo{year}{2016}),
  \bibinfo{pages}{7408--7423}.
\newblock


\bibitem[\protect\citeauthoryear{Arasu, Babu, and Widom}{Arasu
  et~al\mbox{.}}{2006}]%
        {Arasu:2006:CCQ:1146461.1146463}
\bibfield{author}{\bibinfo{person}{Arvind Arasu}, \bibinfo{person}{Shivnath
  Babu}, {and} \bibinfo{person}{Jennifer Widom}.}
  \bibinfo{year}{2006}\natexlab{}.
\newblock \showarticletitle{The CQL Continuous Query Language: Semantic
  Foundations and Query Execution}.
\newblock \bibinfo{journal}{{\em The VLDB Journal\/}} \bibinfo{volume}{{15}, 2}
  (\bibinfo{date}{June} \bibinfo{year}{2006}), \bibinfo{pages}{121--142}.
\newblock
\showISSN{1066-8888}


\bibitem[\protect\citeauthoryear{Bellavista and Zanni}{Bellavista and
  Zanni}{2017}]%
        {Bellavista:2017:FFC:3007748.3007777}
\bibfield{author}{\bibinfo{person}{Paolo Bellavista} {and}
  \bibinfo{person}{Alessandro Zanni}.} \bibinfo{year}{2017}\natexlab{}.
\newblock \showarticletitle{Feasibility of Fog Computing Deployment Based on
  Docker Containerization over RaspberryPi}. In \bibinfo{booktitle}{{\em
  Proceedings of the 18th International Conference on Distributed Computing and
  Networking}} \bibinfo{series}{{\em (ICDCN '17)}}. ACM, Article
  \bibinfo{articleno}{16}, 10 pages.
\newblock
\showISBNx{978-1-4503-4839-3}


\bibitem[\protect\citeauthoryear{Bonomi, Milito, Natarajan, and Zhu}{Bonomi
  et~al\mbox{.}}{2014}]%
        {bonomi2014fog}
\bibfield{author}{\bibinfo{person}{Flavio Bonomi}, \bibinfo{person}{Rodolfo
  Milito}, \bibinfo{person}{Preethi Natarajan}, {and} \bibinfo{person}{Jiang
  Zhu}.} \bibinfo{year}{2014}\natexlab{}.
\newblock \showarticletitle{Fog computing: A platform for internet of things
  and analytics}.
\newblock In \bibinfo{booktitle}{{\em Big Data and Internet of Things: A
  Roadmap for Smart Environments}}. Springer, \bibinfo{pages}{169--186}.
\newblock


\bibitem[\protect\citeauthoryear{Cisco}{Cisco}{2017}]%
        {cisco}
\bibfield{author}{\bibinfo{person}{Cisco}.} \bibinfo{year}{2017}\natexlab{}.
\newblock \bibinfo{title}{{Cisco DevNet: IOx}}.
\newblock \url{https://developer.cisco.com/site/iox/}.
  (\bibinfo{year}{2017}).
\newblock
\newblock
\shownote{[Online; accessed 19-January-2017].}


\bibitem[\protect\citeauthoryear{Cugola and Margara}{Cugola and
  Margara}{2010}]%
        {Cugola:2010:TFD:1827418.1827427}
\bibfield{author}{\bibinfo{person}{Gianpaolo Cugola} {and}
  \bibinfo{person}{Alessandro Margara}.} \bibinfo{year}{2010}\natexlab{}.
\newblock \showarticletitle{TESLA: A Formally Defined Event Specification
  Language}. In \bibinfo{booktitle}{{\em Proceedings of the Fourth ACM
  International Conference on Distributed Event-Based Systems}}
  \bibinfo{series}{{\em (DEBS '10)}}. ACM, \bibinfo{pages}{50--61}.
\newblock
\showISBNx{978-1-60558-927-5}


\bibitem[\protect\citeauthoryear{Eilander, Trambauer, Wagemaker, and van
  Loenen}{Eilander et~al\mbox{.}}{2016}]%
        {eilander2016harvesting}
\bibfield{author}{\bibinfo{person}{Dirk Eilander}, \bibinfo{person}{Patricia
  Trambauer}, \bibinfo{person}{Jurjen Wagemaker}, {and}
  \bibinfo{person}{Arnejan van Loenen}.} \bibinfo{year}{2016}\natexlab{}.
\newblock \showarticletitle{Harvesting social media for generation of near
  real-time flood maps}.
\newblock \bibinfo{journal}{{\em Procedia Engineering\/}}
  \bibinfo{volume}{154} (\bibinfo{year}{2016}), \bibinfo{pages}{176--183}.
\newblock


\bibitem[\protect\citeauthoryear{Gu, Zhou, Fu, and Wan}{Gu
  et~al\mbox{.}}{2015}]%
        {7127659}
\bibfield{author}{\bibinfo{person}{Y. Gu}, \bibinfo{person}{M. Zhou},
  \bibinfo{person}{S. Fu}, {and} \bibinfo{person}{Y. Wan}.}
  \bibinfo{year}{2015}\natexlab{}.
\newblock \showarticletitle{Airborne WiFi networks through directional
  antennae: An experimental study}. In \bibinfo{booktitle}{{\em 2015 IEEE
  Wireless Communications and Networking Conference (WCNC)}}.
  \bibinfo{pages}{1314--1319}.
\newblock
\showISSN{1525-3511}


\bibitem[\protect\citeauthoryear{Hong, Lillethun, Ramachandran,
  Ottenw\"{a}lder, and Koldehofe}{Hong et~al\mbox{.}}{2013}]%
        {Hong:2013:MFP:2491266.2491270}
\bibfield{author}{\bibinfo{person}{Kirak Hong}, \bibinfo{person}{David
  Lillethun}, \bibinfo{person}{Umakishore Ramachandran}, \bibinfo{person}{Beate
  Ottenw\"{a}lder}, {and} \bibinfo{person}{Boris Koldehofe}.}
  \bibinfo{year}{2013}\natexlab{}.
\newblock \showarticletitle{Mobile Fog: A Programming Model for Large-scale
  Applications on the Internet of Things}. In \bibinfo{booktitle}{{\em
  Proceedings of the Second ACM SIGCOMM Workshop on Mobile Cloud Computing}}
  \bibinfo{series}{{\em (MCC '13)}}. ACM, \bibinfo{pages}{15--20}.
\newblock
\showISBNx{978-1-4503-2180-8}


\bibitem[\protect\citeauthoryear{Liu, Liu, Gao, Gong, Kang, Zhi, Chi, and
  Shi}{Liu et~al\mbox{.}}{2015}]%
        {doi:10.1080/00045608.2015.1018773}
\bibfield{author}{\bibinfo{person}{Yu Liu}, \bibinfo{person}{Xi Liu},
  \bibinfo{person}{Song Gao}, \bibinfo{person}{Li Gong},
  \bibinfo{person}{Chaogui Kang}, \bibinfo{person}{Ye Zhi},
  \bibinfo{person}{Guanghua Chi}, {and} \bibinfo{person}{Li Shi}.}
  \bibinfo{year}{2015}\natexlab{}.
\newblock \showarticletitle{Social Sensing: A New Approach to Understanding Our
  Socioeconomic Environments}.
\newblock \bibinfo{journal}{{\em Annals of the Association of American
  Geographers\/}} \bibinfo{volume}{{105}, 3} (\bibinfo{year}{2015}),
  \bibinfo{pages}{512--530}.
\newblock


\bibitem[\protect\citeauthoryear{Lochert, Hartenstein, Tian, Fussler, Hermann,
  and Mauve}{Lochert et~al\mbox{.}}{2003}]%
        {lochert2003routing}
\bibfield{author}{\bibinfo{person}{Christian Lochert}, \bibinfo{person}{Hannes
  Hartenstein}, \bibinfo{person}{Jing Tian}, \bibinfo{person}{Holger Fussler},
  \bibinfo{person}{Dagmar Hermann}, {and} \bibinfo{person}{Martin Mauve}.}
  \bibinfo{year}{2003}\natexlab{}.
\newblock \showarticletitle{A routing strategy for vehicular ad hoc networks in
  city environments}. In \bibinfo{booktitle}{{\em Intelligent Vehicles
  Symposium, 2003. Proceedings. IEEE}}. IEEE, \bibinfo{pages}{156--161}.
\newblock


\bibitem[\protect\citeauthoryear{Manoj and Baker}{Manoj and Baker}{2007}]%
        {Manoj:2007:CCE:1226736.1226765}
\bibfield{author}{\bibinfo{person}{B.S. Manoj} {and}
  \bibinfo{person}{Alexandra~Hubenko Baker}.} \bibinfo{year}{2007}\natexlab{}.
\newblock \showarticletitle{Communication Challenges in Emergency Response}.
\newblock \bibinfo{journal}{{\em Commun. ACM\/}} \bibinfo{volume}{{50}, 3}
  (\bibinfo{date}{March} \bibinfo{year}{2007}), \bibinfo{pages}{51--53}.
\newblock
\showISSN{0001-0782}


\bibitem[\protect\citeauthoryear{Mayer, Koldehofe, and Rothermel}{Mayer
  et~al\mbox{.}}{2015}]%
        {7024105}
\bibfield{author}{\bibinfo{person}{Ruben Mayer}, \bibinfo{person}{Boris
  Koldehofe}, {and} \bibinfo{person}{Kurt Rothermel}.}
  \bibinfo{year}{2015}\natexlab{}.
\newblock \showarticletitle{Predictable Low-Latency Event Detection With
  Parallel Complex Event Processing}.
\newblock \bibinfo{journal}{{\em IEEE Internet of Things Journal\/}}
  \bibinfo{volume}{{2}, 4} (\bibinfo{date}{Aug} \bibinfo{year}{2015}),
  \bibinfo{pages}{274--286}.
\newblock


\bibitem[\protect\citeauthoryear{Merchant, Elmer, and Lurie}{Merchant
  et~al\mbox{.}}{2011}]%
        {merchant2011integrating}
\bibfield{author}{\bibinfo{person}{Raina~M Merchant}, \bibinfo{person}{Stacy
  Elmer}, {and} \bibinfo{person}{Nicole Lurie}.}
  \bibinfo{year}{2011}\natexlab{}.
\newblock \showarticletitle{Integrating social media into
  emergency-preparedness efforts}.
\newblock \bibinfo{journal}{{\em New England Journal of Medicine\/}}
  \bibinfo{volume}{{365}, 4} (\bibinfo{year}{2011}), \bibinfo{pages}{289--291}.
\newblock


\bibitem[\protect\citeauthoryear{Pentland, Fletcher, and Hasson}{Pentland
  et~al\mbox{.}}{2004}]%
        {1319279}
\bibfield{author}{\bibinfo{person}{A. Pentland}, \bibinfo{person}{R. Fletcher},
  {and} \bibinfo{person}{A. Hasson}.} \bibinfo{year}{2004}\natexlab{}.
\newblock \showarticletitle{DakNet: rethinking connectivity in developing
  nations}.
\newblock \bibinfo{journal}{{\em Computer\/}} \bibinfo{volume}{{37}, 1}
  (\bibinfo{date}{Jan} \bibinfo{year}{2004}), \bibinfo{pages}{78--83}.
\newblock
\showISSN{0018-9162}


\bibitem[\protect\citeauthoryear{Simon, Goldberg, and Adini}{Simon
  et~al\mbox{.}}{2015}]%
        {Simon2015609}
\bibfield{author}{\bibinfo{person}{Tomer Simon}, \bibinfo{person}{Avishay
  Goldberg}, {and} \bibinfo{person}{Bruria Adini}.}
  \bibinfo{year}{2015}\natexlab{}.
\newblock \showarticletitle{Socializing in emergencies - A review of the use of
  social media in emergency situations}.
\newblock \bibinfo{journal}{{\em International Journal of Information
  Management\/}} \bibinfo{volume}{{35}, 5} (\bibinfo{year}{2015}),
  \bibinfo{pages}{609 -- 619}.
\newblock
\showISSN{0268-4012}


\bibitem[\protect\citeauthoryear{Skjegstad, Johnsen, Bloebaum, and
  Maseng}{Skjegstad et~al\mbox{.}}{2012}]%
        {skjegstad2012mist}
\bibfield{author}{\bibinfo{person}{Magnus Skjegstad}, \bibinfo{person}{Frank~T
  Johnsen}, \bibinfo{person}{Trude~H Bloebaum}, {and} \bibinfo{person}{Torleiv
  Maseng}.} \bibinfo{year}{2012}\natexlab{}.
\newblock \showarticletitle{Mist: A reliable and delay-tolerant
  publish/subscribe solution for dynamic networks}. In \bibinfo{booktitle}{{\em
  New Technologies, Mobility and Security (NTMS), 2012 5th International
  Conference on}}. IEEE, \bibinfo{pages}{1--8}.
\newblock


\bibitem[\protect\citeauthoryear{Stiegler, Tilley, and Parveen}{Stiegler
  et~al\mbox{.}}{2011}]%
        {6081815}
\bibfield{author}{\bibinfo{person}{R. Stiegler}, \bibinfo{person}{S. Tilley},
  {and} \bibinfo{person}{T. Parveen}.} \bibinfo{year}{2011}\natexlab{}.
\newblock \showarticletitle{Finding family and friends in the aftermath of a
  disaster using federated queries on social networks and websites}. In
  \bibinfo{booktitle}{{\em 2011 13th IEEE Int'l Symposium on Web Systems
  Evolution (WSE)}}. \bibinfo{pages}{21--26}.
\newblock
\showISSN{1550-4441}


\bibitem[\protect\citeauthoryear{Yusuf and Ramachandran}{Yusuf and
  Ramachandran}{2012}]%
        {lateef2012}
\bibfield{author}{\bibinfo{person}{L. Yusuf} {and} \bibinfo{person}{U.
  Ramachandran}.} \bibinfo{year}{2012}\natexlab{}.
\newblock \showarticletitle{Community Membership Management for Transient
  Social Networks}. In \bibinfo{booktitle}{{\em 2012 21st International
  Conference on Computer Communications and Networks (ICCCN)}}.
  \bibinfo{pages}{1--7}.
\newblock


\bibitem[\protect\citeauthoryear{İlker Bekmezci, Sahingoz, and Şamil
  Temel}{İlker Bekmezci et~al\mbox{.}}{2013}]%
        {Bekmezci20131254}
\bibfield{author}{\bibinfo{person}{İlker Bekmezci},
  \bibinfo{person}{Ozgur~Koray Sahingoz}, {and} \bibinfo{person}{Şamil
  Temel}.} \bibinfo{year}{2013}\natexlab{}.
\newblock \showarticletitle{Flying Ad-Hoc Networks (FANETs): A survey}.
\newblock \bibinfo{journal}{{\em Ad Hoc Networks\/}} \bibinfo{volume}{{11}, 3}
  (\bibinfo{year}{2013}), \bibinfo{pages}{1254 -- 1270}.
\newblock
\showISSN{1570-8705}


\end{thebibliography}
